\documentstyle[aps,epsfig,amsfonts]{revtex}

\newcommand \bea {\begin{eqnarray} \nonumber }
\newcommand \ee {\end{equation}}
\newcommand \eea {\end{eqnarray}}
\newcommand \be {\begin{equation}}

\begin{document}
%\jl{1}

\title{Basins of attraction of metastable states of the spherical
$p$-spin model}
\author{A. Barrat, S. Franz}

\address{ International Center for Theoretical Physics\\
Strada Costiera 11,
P.O. Box 563,
34100 Trieste (Italy)}

\date{\today}

\maketitle

\begin{abstract}
We study  the basins of attraction of metastable 
states in the spherical $p$-spin spin glass
model, starting the relaxation dynamics 
at a given distance from a thermalized condition. 
Weighting the initial condition with the Boltzmann 
distribution we find a finite size for the basins.
On the contrary, a white weighting of 
the initial condition implies  vanishing basins of 
attraction. 
We make the corresponding of our results with the 
ones of a recently constructed effective potential. 
\end{abstract}

\vskip .5cm
%\newpage

The so-called spherical $p$-spin spin glass model has been the 
subject of many studies: indeed, being mean-field, it allows for a 
detailed analytic study, while still displaying very rich static
and dynamical behaviours. 
In paricular in recent papers \cite{I,cavagna1,cavagna2},
it has been shown that the structure of its metastable
states, which dominate the Gibbs measure between two temperatures
noted $T_s$ and $T_d$ (where the static and dynamic transitions
respectively occur \cite{crihorsom}),
is very rich and complex. The existence of these
states is revealed using the approach of Thouless, Anderson and Palmer
\cite{TAP} and they are therefore often called ``TAP states''
\cite{kuparvir,crisomtap}.
These recent works make use of the real replicas method: copies
of the system are considered, at various distances from each
other, and the free energy cost (called ''effective potential'' function)
to keep them at given distances is computed. The minima of the potential
can then be associated with the fact that the replicas lie in metastable 
states, thus give informations on the distances between states.

This method is therefore purely static; indeed, the dynamics after a quench
do not see at all the metastable states, giving instead rise to
the phenomenon known as aging \cite{cuku}: quenched below $T_d$,
the system remains out of equilibrium for all times with an energy 
higher then the thermodynamic one.
On the other hand, it was shown in \cite{I,babumez,bfp} that particular initial
conditions for the dynamics (namely, taking the system thermalized at a certain
temperature between $T_s$ and $T_d$, and then letting it evolve at another
temperature) could allow a dynamical exploration of the metastable states
finding results consistent with the picture coming from the 
two replica potential. 

In this letter we address the problem of determining the size of 
the basins of attraction of the TAP states. This will be done studying the
Langevin relaxation of a system starting at an initial time at a given 
fixed overlap ($q_{12}$ in the following) from an equilibrium configuration. 
This, of course, does not specify completely the 
initial conditions. In the following we will
consider two families of them
(for fixed $q_{12}$), weighting 
the initial conditions with the Boltzmann distribution and 
with the uniform one. 

As we will see, the case of Boltzmann weighting can be related 
to the results found for a three-replica potential \cite{cavagna1}
recently introduced. 
Therefore, for future use,  we  briefly recall the results of the two-
and three-replicas potentials, $V_2$ and $V_3$, 
for the $p$-spin model \footnote{we will not recall the details of
the computations, which can be found in \cite{cavagna1}}:
indeed, the results of the study of $V_2$ will be used
to determine the initial conditions of the dynamics, and those of
the study of $V_3$ will be compared to the outcome of the dynamics.

The two-replicas potential $V_2$ 
is defined \cite{I} as the free energy cost to
keep a configuration $\tau$ at a fixed overlap $q_{12}$ with an equilibrium
configuration $\sigma$. While in general the two replicas
can be at different temperatures, we will here limit ourselves
to the case of equal temperature $T$ for replicas 1 ($\sigma$)
and 2 ($\tau$).
The overlap between the replicas is denoted $q_{12}$, while the use
of the replica trick leads to a description of the second replica
by a one step RSB matrix $Q^{22}$ of parameters $(r_1,r_0,x)$, determined
variationally. 
The absolute minimum of the potential is always for
$q_{12}=0$: then the second
replica is at equilibrium, with no constraint,
so the free energy cost is zero. For
$T < T_d$, $\sigma$ lies in one of the metastable states that dominate
the statics (of Edwards-Anderson parameter $q_{EA}$),
and a relative minimum appears for a non-zero 
value of $q_{12}$: it corresponds to having the second  replica
in the same TAP state as the first.

In order to study the organization of the metastable states in the phase
space, the construction was generalized to three replicas
in \cite{cavagna1}. There, a first replica $\rho$ is free to thermalize
at $T$; a second replica $\sigma$ is constrained to thermalize
at $T$ with a fixed overlap $q_{12}$ with $\rho$, and the potential
$V_3$ is defined as the free energy cost to keep a third replica
$\tau$ (at the same temperature $T$)
at overlaps $q_{13}$ from $\rho$ and $q_{23}$ from $\sigma$. We will
take $T_s < T < T_d$: then the first replica is in a certain TAP
state of equilibrium at $T$.

Since the two first replicas are independent from the third, the
overlap matrices that describe them are identical to the ones used
for $V_2$. The third replica is described by a one step RSB matrix.
In the minima of the potential, this matrix is in fact replica
symmetric, with only one parameter $q_{33}$.

\begin{figure}
\centerline{\hbox{
\epsfig{figure=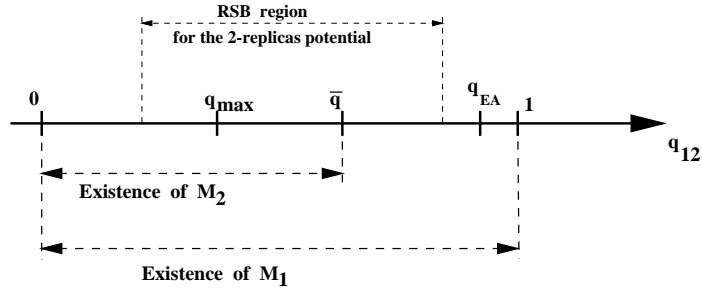,width=4cm,angle=-90}
}}
\caption{Domains of existence of the minima $M_1$ and $M_2$ of the 
potential $V_3$}
\label{q12}
\end{figure}

The analysis of \cite{cavagna1} showed that, depending on the
value of $q_{12}$, the potential can have one or two non trivial
minima (apart from the minimum at $q_{13}=q_{23}=0$ corresponding
to the third replica in an unspecified equilibrium
state at $T$, different from the ones of replicas
1 and 2).
The first minimum, called $M_1$, exists for any value of $q_{12}$, and
has $q_{13}=q_{EA}$ and $q_{23} \approx q_{12}$.
Its interpretation is that the third replica lies in the same state
as the first one. It therefore exists independently from the value of
$q_{12}$.
The second, more interesting minimum, called $M_2$ in \cite{cavagna1},
corresponds to the third replica close to the second one
($q_{23} \approx q_{EA}$, $q_{13} \approx q_{12} < q_{EA}$).
Its interpretation is that
the second replica lies in the basin of attracion of 
a metastable state at non-zero overlap
$q_{12}$ from the first replica, while the third replica is at
equilibrium in this state.
This solution exists only for values of $q_{12}$ lower than a certain
$\bar{q}$ (which depends on the temperature), with $\bar{q} < q_{EA}$
(see figure (\ref{q12})).
This $\bar{q}$ gives therefore the minimum distance (or maximum
overlap), from an equilibrium
state at $T$, at which can be found another metastable state.

In order to study the basins of attraction of the metastable 
states, we study the relaxation of a system with the following 
initial condition.
We consider a reference configuration 
at equilibrium at temperature $T$. Then 
the system evolves from a configuration thermalized 
at temperature $T$, but with the constraint that its overlap
with the reference configuration is equal to $q_{12}$. 
At positive time the spins evolve according to an (unconstrained) Langevin
dynamics at temperature $T$:
\be
\frac{d \sigma_i(t)}{dt} = -\frac{\partial H}{\partial \sigma_i} 
- \mu(t) \sigma_i(t)
 + \eta_i(t),
\label{langepspin}
\ee
where the $\eta_i$ are Gaussian thermal noises with
$\langle \eta_i(t) \eta_j(t') \rangle = 2T\delta_{ij}
\delta(t-t') $, and $\mu(t)$ is a multiplier 
that implements the spherical constraint 
$\sum_{i} \sigma_i^2 = N$ at all times.

The goal will be to see how the system evolves dynamically, depending on
the value of the initial overlap with an equilibrium configuration.

In order to implement the initial conditions, we have to use the 
replica trick to describe the systems: the first one will be described
by  $\rho^a$ with $a=1,\cdots,n$; the initial conditions are
$\sigma^\alpha$, $\alpha = 1, \cdots,m$ and only $\sigma^1$ is evolving
with time, so we will note it $\sigma(t)$ instead of $\sigma^1(t)$, with
$\sigma(0)=\sigma^1(0)$. In the end the limits $m \to 0$, $n \to 0$ are
taken, with a one-step replica symmetric breaking Ansatz.
In the infinite $N$ limit,
we can obtain, in the same way as in \cite{houghton,kt,I,babumez},
a set of coupled self consistent
dynamical equations for the following quantities (the equations
are written in appendix):
\bea
C(t,t')&=&\frac{1}{N} 
\sum_{i=1}^N\overline{\langle  \sigma_i(t) \sigma_i(t') \rangle}
\ ;\ 
R(t,t') = \frac{1}{N} \sum_i
\overline{\langle \frac{\partial \sigma_i(t)}{\partial \eta_i(t')} \rangle}
\nonumber \\
C_\alpha(t) &=&\frac{1}{N} 
\sum_{i=1}^N \overline{\langle \sigma_i^\alpha \sigma_i(t) \rangle}
\ \  (\alpha > 1\ ; \mbox{case}\ 
\alpha =1\ :\ C_1(t)=C(t,0)) 
\ ;\ 
Q_a(t)=\frac{1}{N} 
\sum_{i=1}^N \overline{\langle \sigma_i(t) \rho_i^a \rangle}
\eea

While $C(t,t')$ and $R(t,t')$ are the usual correlation and response functions
of the system, the evolution of the $C_\alpha$ and $Q_a$ 
will give informations on how the system departs from its initial conditions
and how close it goes to the equilibrium state of the first real
replica $\rho$.

The initial conditions can be obtained from the study of the
two-replicas potential: we first impose
$Q_a(0)=Q^{12}_{a,1} = \delta_{a,1} q_{12}$. Since 
the structure of the equations respect the replica symmetric character
or the breaking of replica symmetry of the $C_\alpha$ and $Q_a$, 
at all times $Q_a(t)=  \delta_{a,1} Q(t)$, with
$Q(t)= \frac{1}{N} \sum_{i=1}^N  \rho_i \sigma_i(t)$. Then, from 
the value of $q_{12}$, we use the two replicas potential to
deduct the values for $C_\alpha(0)$.
According to the value of $q_{12}$, it can be
replica symmetric or have one step of replica symmetry breaking.
Since the replica symmetric case can be recovered in a simple way
from the equations
of the RSB case, we will consider only the one step case.
Then the initial conditions for $C_\alpha$ has parameters $(r_1,r_0,x)$;
therefore at all times the $C_\alpha(t)$ will have the form
\be
C_\alpha(t)=(C(t,0),C_1(t),C_0(t))
\ee
with the same breaking parameter $x$
\footnote{This means that $C_\alpha (t)$ is equal to
$C(t,0)$ for $\alpha=1$, to $C_1(t)$ for $\alpha=2,...,x$, and to
$C_0(t)$ for $\alpha=x+1,...,n$.}, and
$C_1(0)=r_1$, $C_0(0)=r_0$
\footnote{to recover the RS case, we take $r_1=r_0$: then,
at all times, $C_1(t)=C_0(t)$.}.
We define the following limiting values:
\bea
\lim_{t \to \infty} Q(t)=q_0\ \  ;\ 
\lim_{t \to \infty} C(t,0) = \tilde{p} \nonumber \\
\lim_{t \to \infty} C_1(t)=c_1 \ \  ;\ 
\lim_{t \to \infty} C_0(t)=c_0 \ \ ;\ 
\lim_{t \to \infty} \mu(t) = \mu
\eea

A simple check is to look at what happens in two extreme cases:
(i) $q_{12}=1$: the system starts at equilibrium at $T$;
then $r_1=r_0=1$, and obviously we obtain
$C(t,0)=Q(t)=C_1(t)=C_0(t)$, $q_0=\tilde{p}=c_1=c_0=q_{EA}$: the
system thermalizes in the particular equilibrium state at $T$ chosen
by $\rho$;
(ii) $q_{12}=0$: the system is not constrained, so clearly we
obtain $Q(t)=C_1(t)=C_0(t)=0$, $C(t,t')=C(t-t')=C(\tau)$: the
system thermalizes in an unspecified TAP state of equilibrium at $T$
\cite{I,babumez}. 

For other values of $q_{12}$,
the numerical integration of the dynamical equations shows that
the system, after a transient, reaches
a certain equilibrium behaviour; to study the system at long times,
we therefore make the Ansatz:
$C(t,t')=C(t-t')=C(\tau)$, 
$R(t,t')=R(t-t')=R(\tau)=- \frac{1}{T} \frac{d C}{d\tau}$,
with
$\lim_{\tau \to \infty} C(\tau) = q$.
This Ansatz allows, with usual methods, 
to obtain coupled equations for the limiting values
of the various one-time quantities, and moreover
we have for the evolution of the asymptotic correlation function:
\be
\frac{d C}{d \tau} = -\frac{T}{1-q} (C(\tau) - q)
-\beta \int_0^\tau du\ (f'(C(\tau -u)) - f'(q)) C'(u).
\label{relax}
\ee

The equations for the values of
$q$, $q_0$, $\tilde{p}$, $c_1$, $c_0$ are equivalent to the equations 
for the parameters $q_{33}$, $q_{13}$, $q_{23}$, $w_{23}$, $z_{23}$ of
the three-replicas potential
\footnote{in the RS case, the equivalence is $q_{33}= q$,
$q_0=q_{13}$, $c_1=c_0=\tilde{p}=q_{23}$. }. This correspondence
pushes forward the one noted in \cite{I,babumez} between the 
potential with two replicas and the dynamics with thermalized initial
conditions.
Here, the interpretation is that
the first replica $\rho$ lies in an equilibrium state at temperature $T$,
the replica number 2 of the potential gives the initial conditions
of the dynamics, while the dynamical system goes towards a minimum
of the potential (given in the potential approach by the
third replica), where it relaxes according to (\ref{relax}), which is exactly
the equation of the relaxation in a TAP state of self-overlap $q$, as given
in \cite{babumez}. 

An important difference between the two approaches is that,
while the potential can be explored for all values of $q_{12}$, $q_{13}$,
$q_{23}$ (i.e. all positions of both replicas 2 and 3),
the only parameter of the studied dynamics is $q_{12}$: the
initial conditions of the dynamics can be compared to the second replica
of the potential, and all possible values of $q_{12}$ can be studied,
but the values of $q_0$, $\tilde{p}$ are {\it outcomes} of the dynamics
and are not chosen. From the equivalence between potential
and dynamics, it follows for the dynamics
that, while for $q_{12}>\bar{q}$ only the solution 
$M_1$ exists, for $q_{12}<\bar{q}$ there are the two solutions
$M_1$ and $M_2$. However, as the dynamical equations 
admit (within the 1-step RSB Ansatz we use) 
a unique solution for any finite time, only one of the two 
can be reached dynamically. The size of the basin of attraction 
of the equilibrium states is related to the smallest value of 
$q_{12}$ for which the solution $M_1$ is reached. 

In order to settle this question we
integrate numerically the dynamical equations (\ref{A2}).
For simplicity we limit our analysis to the case $p=3$ for
which most of the analysis of \cite{cavagna1} was done. 
Then $T_s \approx 0.586$, $T_d \approx 0.61237$ and we will show results for 
$T=0.6$: then $q_{EA}=0.6$, $\bar{q} \approx 0.342 $.
The numerical integration is done using a simple iteration algorithm, 
discretizing the dynamical equations with a finite time-step $h$. We proceed
with three values $h$, $2h$, $4h$ and then do the interpolation
at $h \to 0$ to compare the numerical values with the values obtained
from the study of the potential $V_3$
\footnote{ after checking that the interpolation at $h \to 0$
coincides well with the values obtained via the potential, we
used a unique value of the time step for some runs that involved
larger timescales, at values of $q_{12}$ close to the
limit of the attraction basin.}.

\begin{figure}
\centerline{\hbox{
\epsfig{figure=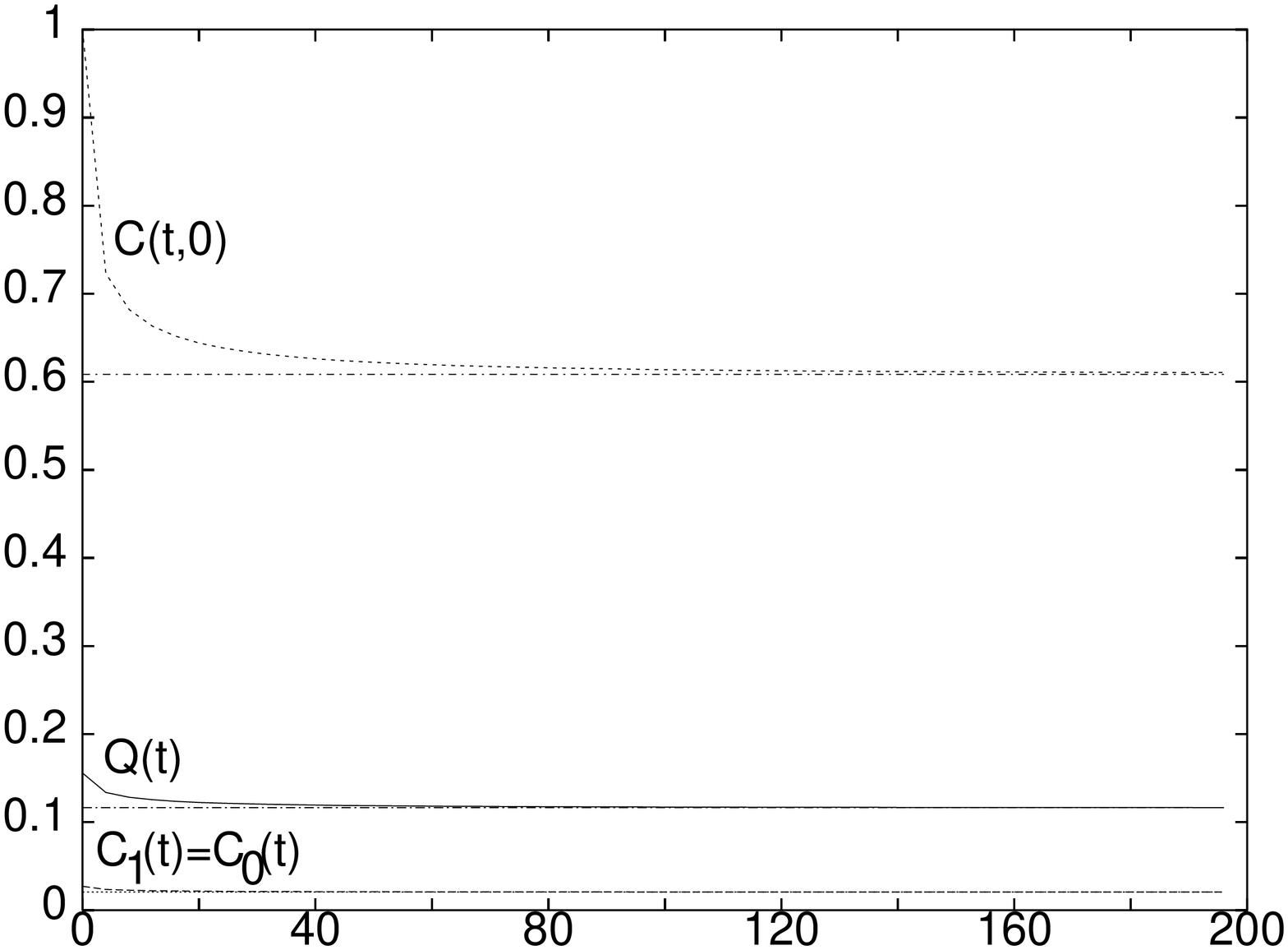,width=5.5cm,angle=-90}
\epsfig{figure=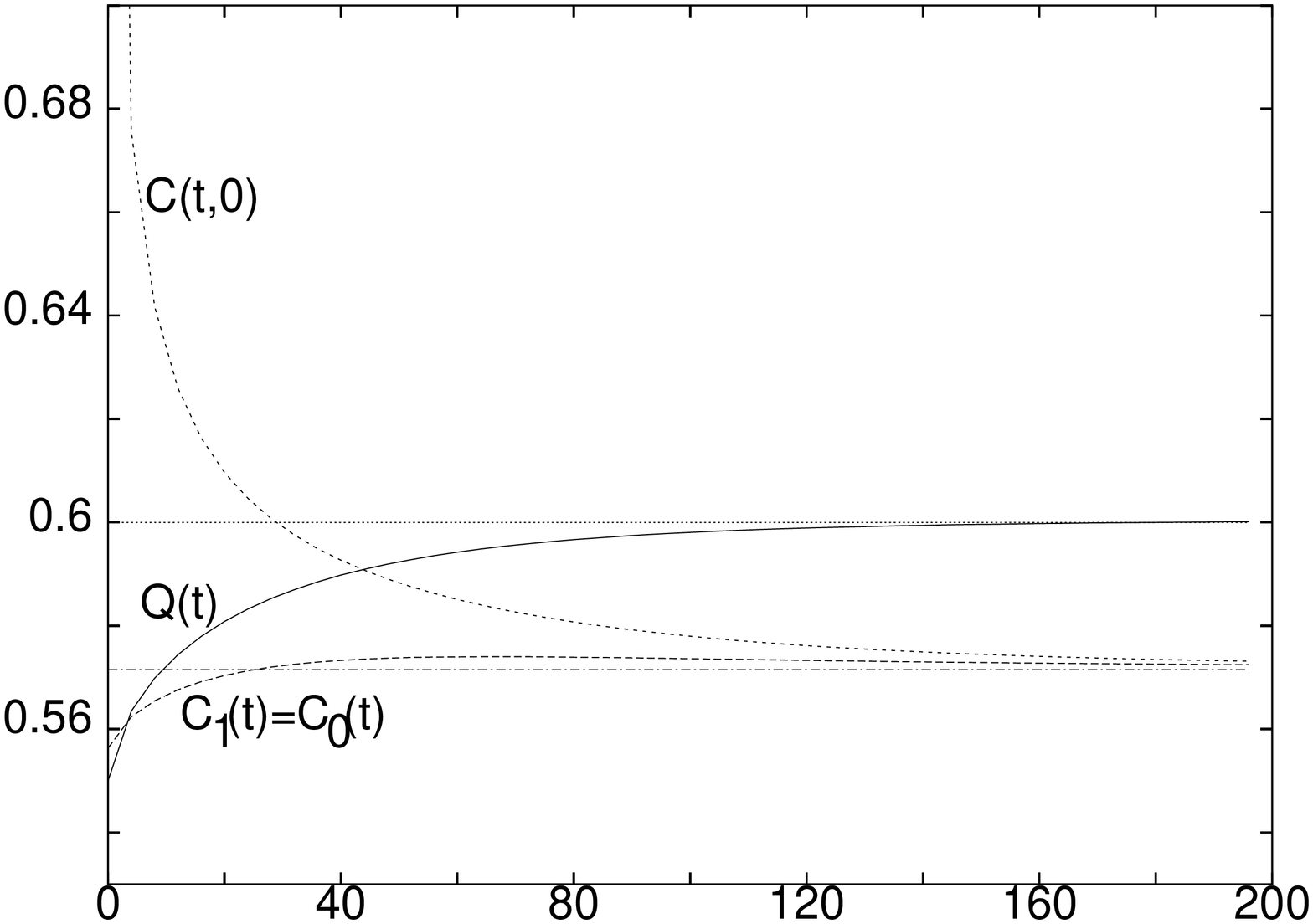,width=5.5cm,angle=-90}
}}
\caption{Evolution of $C(t,0)$, $Q(t)$ and $C_1(t)=C_0(t)$ with time for 
$q_{12}=0.155702$ (left) and $q_{12}=0.55$ (right). These curves 
correspond to the extrapolation at $h \to 0$ of the results of the
numerical integration with $h=0.05,\ 0.1,\ 0.2$;
we see that they go quite fast to their limiting 
values $\tilde{p}$, $q_0$ and $c_1=c_0$ (given by the horizontal
lines), with moreover $\tilde{p}=c_1=c_0$ for $q_{12}=0.55$.}
\label{q12.15.55}
\end{figure}

For ``small'' values of $q_{12}$, the dynamics converge rapidly
towards an equilibrium behaviour with time tranlsation 
invariance and fluctuation-dissipation relation. 
The numerical integration yields
limits in excellent accordance with the 
resolution of the aforementioned equations for 
$q$, $q_0$, $\tilde{p}$, $c_1$, $c_0$,
and coincide with the values of the various parameters in the 
minimum $M_2$ of the three replicas potential.
(e.g., see figure (\ref{q12.15.55}), $q_{12}=0.155702$, then $r_1=r_0=0.02715$,
$q_0=0.116338$, $c_1=c_0=0.0205437$, 
$q=0.608423$, $\tilde{p}=0.609467$. These values coincide with
the values respectively of $q_{13}$, $w_{23}=z_{23}$, $q_{33}$,
$q_{23}$  in the minimum $M_2$.)
This means that the system $\sigma(t)$ stays in the state found by the replica 
$\sigma(0)$, with finite overlap with $\rho$. 
We are therefore in the minimum $M_2$, and out of the attraction basin
of the equilibrium state where $\rho$ lies.

For $q_{12}$ ``large'', conversely,  we expect that the system, which
starts close to $\rho$, remains in the same state. This is indeed what we find
(e.g. for $q_{12}=0.817272$, then $r_1=r_0=0.7859$; we obtain
$q_0=q=q_{EA}$, $c_1=c_0=\tilde{p}=0.633625$;
the integration of two-times equations coincide well with
the integrations of the equation on $C(\tau)$ and with these values).
For a somewhat smaller value of $q_{12}$ (e.g., see figure (\ref{q12.15.55}),
$q_{12}=0.55$,  $r_1=r_0=0.556345$), 
the same behaviour is obtained:
the two times equations yield the same results as the equations
using the equilibrium 
Ansatz, with $q_0=q=q_{EA}$, $c_1=c_0=\tilde{p}=0.571534$).

For these values of $q_{12}$, the system thermalizes therefore in the
TAP state found by $\rho$. The long time dynamics is the relaxation
dynamics in a 
TAP state of equlibrium at $T$, therefore do not depend
on the initial conditions.

\begin{figure}
\centerline{\hbox{
\epsfig{figure=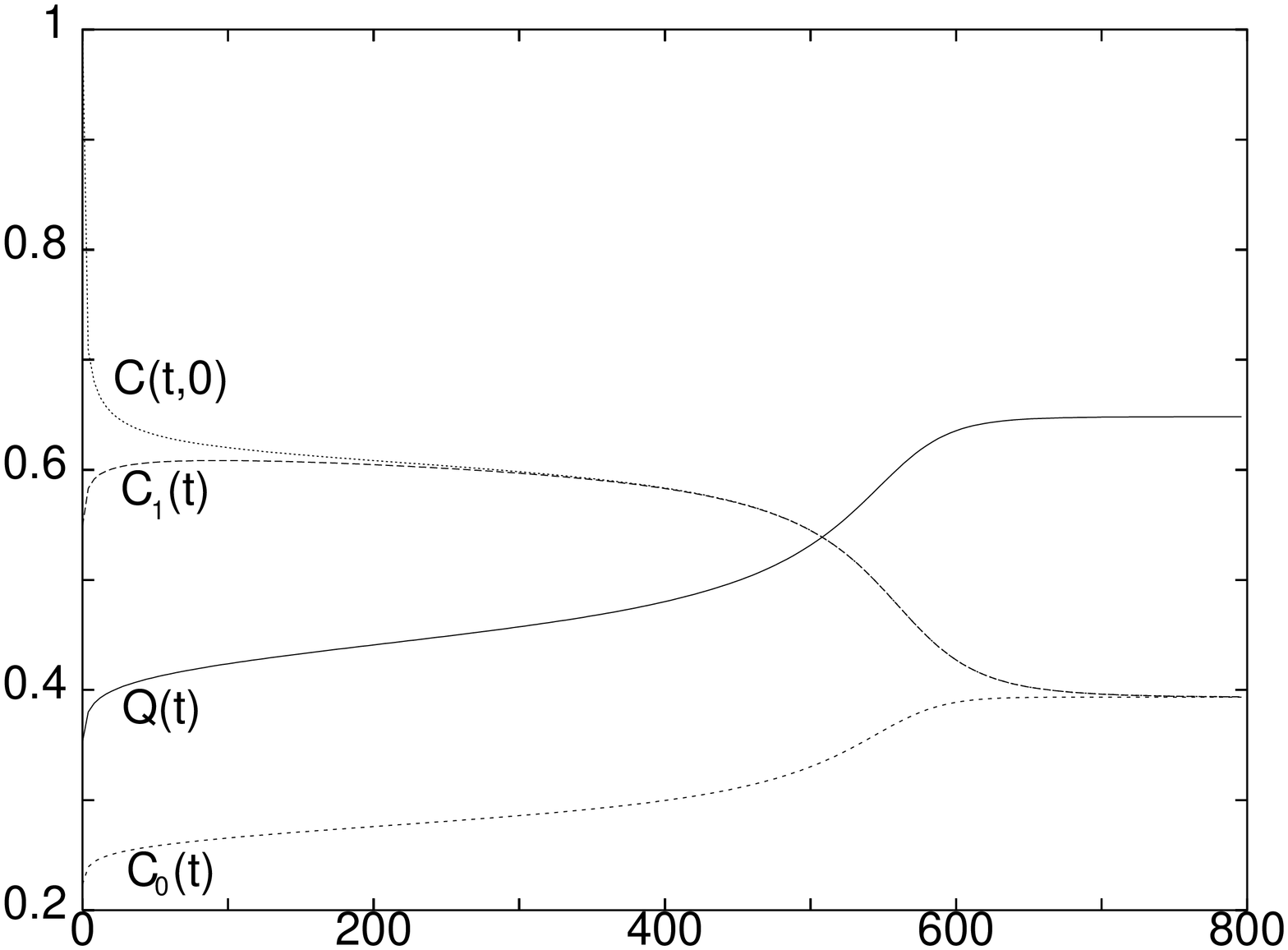,width=5cm,angle=-90}
\epsfig{figure=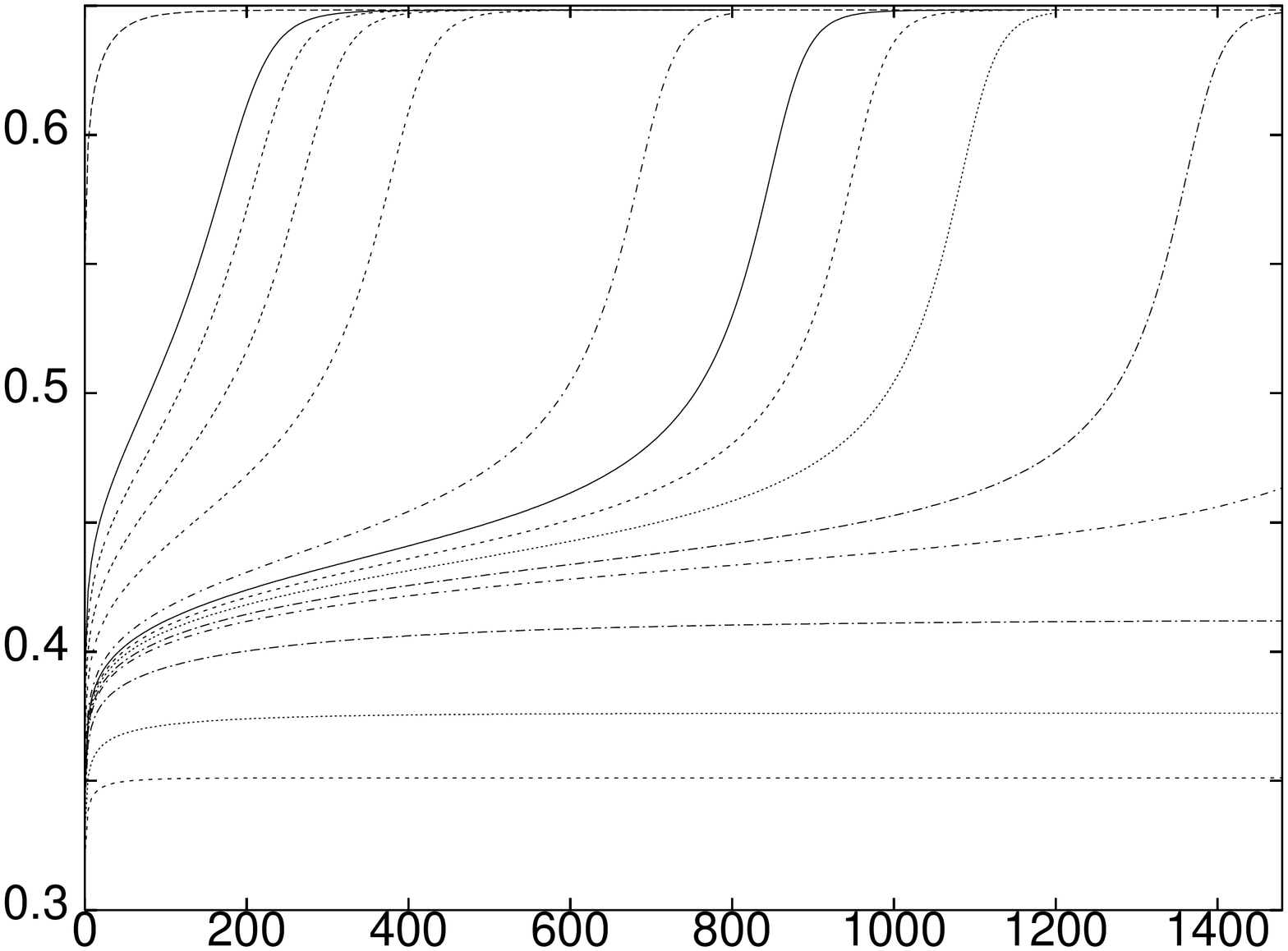,width=5cm,angle=-90}
}}
\caption{Left: Evolution with time 
of the one-time quantities $C(t,0)$, $C_1(t)$,
$C_0(t)$, $Q(t)$, for $q_{12} = 0.353$. We observe the presence of a 
plateau until $t^* \approx 500$ before the quantities reach their final
values corresponding to $M_1$.
Right: Evolution with time of $Q(t)$
for various values of $q_{12}$. From bottom
to top, $q_{12}=\ 0.32,\ 0.33,\ 0.34$ ($< \bar{q}$), and
$q_{12}=0.344,\ 0.345,\ 0.346,\ 0.347,\ 0.348,\ 0.35,\ 0.36,\ 0.37,\ 0.38,\ 
0.39,\ 0.55$ ($> \bar{q}$): we see the
growth of the timescale given by the length of the plateau as
$q_{12}$ decreases towards $\bar{q}$. For $q_{12} > \bar{q}$, we see that
all the curves go to the same limit corresponding to $M_1$, while the limiting
value depends on $q_{12}$ for $q_{12} < \bar{q}$.
In both figures, the time-step used is $h=0.2$.}
\label{q12.353_plateau}
\end{figure}

\begin{figure}
\centerline{\hbox{
\epsfig{figure=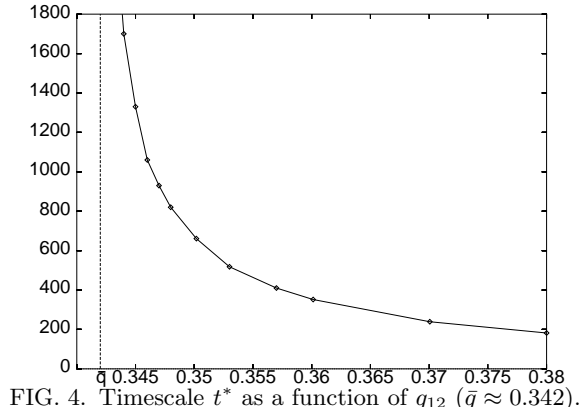,width=5cm,angle=-90}
}}
\caption{Timescale $t^*$ as a function of $q_{12}$ ($\bar{q} \approx 0.342$).}
\label{tstar}
\end{figure}

The outcome of the dynamics for small and large values of $q_{12}$ show
that both minima $M_1$ and $M_2$
can be dynamically reached. We can then naturally ask
if a minimum is always reached, and what is the limiting value of
$q_{12}$ for which the system goes towards $\rho$.
We therefore study the behaviour of the system for values of $q_{12}$ 
decreasing towards $\bar{q}$, and also for values
lower than, but close to $\bar{q}$.
We observe that, while for large values of $q_{12}$ 
(or also for values of $q_{12}$ lower than $\bar{q}$) the one-time
quantities go directly to their limiting values, upon
decreasing $q_{12}$ a plateau appears at an intermediate value 
between the initial and the limiting values. This plateau gives
a timescale $t^*$ that grows and diverges when $q_{12} \to \bar{q}$. We
show the various one-time quantities 
for a particular value of $q_{12}$, and
the evolution of the plateau and of the timescale with $q_{12}$, in
figures (\ref{q12.353_plateau}) and (\ref{tstar})).
For $q_{12}$ lower than $\bar{q}$, we observe that
the dynamics converges towards the values of the parameters
in the minimum $M_2$ of the potential, and do no more reach $M_1$.

The situation is therefore that, for any $q_{12} > \bar{q}$, 
the system reaches
the state where the first replica, $\rho$, lies, but after a transient
which length diverges as $q_{12}$ goes to
$\bar{q}$. For initial conditions farther from $\rho$
than $\bar{q}$, i.e. as soon as the minimum $M_2$ of the potential 
exists, the system relaxes in the state corresponding to $M_2$, which
is a metastable state at finite overlap with $\rho$, and is no more
able to ``reach'' $\rho$. We can therefore understand $\bar{q}$
as the limit of the attraction basin of the state where $\rho$ lies.
It is worth noting that $\bar{q}$ is quite small ($\bar{q} \approx 0.342$ 
for $T=0.6$), which means that it is possible to find configurations
that will dynamically evolve towards a TAP state, and thermalize in it
within a finite time, even at quite large
distances from typical configurations of this state.

\begin{figure}
\centerline{\hbox{
\epsfig{figure=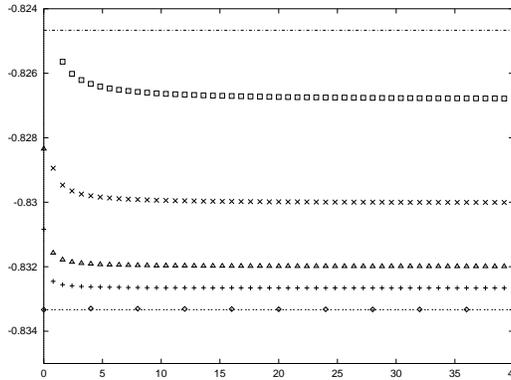,width=5cm,angle=-90}
}}
\caption{Evolution of the energy of the system starting at overlap
$q_{12}=1.,\ 0.999,\ 0.998,\ 0.995,\ 0.99$ (symbols, from bottom to top)
from an equilibrium configuration $\rho$ at $T=0.6$ ($p=3$),
but otherwise randomly; the lines give
the equilibrium energy $E_{eq} \approx -0.83333$ and
$E_{dyn} \approx -0.82467$.}
\label{energies}
\end{figure}

This is the situation that we find if we weight the initial condition
with the Boltzmann probability. Let us turn now briefly 
to the case of initial conditions with overlap $q_{12}$ with $\rho$, but
otherwise uniformly distributed. In this case the basins of attraction
are vanishing. Indeed for any value of $q_{12}$ we find that the 
asymptotic value of the energy is larger than the equilibrium value $E_{eq}$
(the energy of $\rho$). 
As far as the correlations with the initial state are concerned we 
have found two different regimes, separated by a (rather large)
threshold value $q^*$ of 
$q_{12}$ (e.g. $q^* \approx 0.99$ for
$p=3$, $T=0.6$). For $q>q^*$ the system reaches a time translation 
invariant situation, with final energy that depends continuously 
on $q_{12}$ (see fig. (\ref{energies})). The overlap with the initial 
condition tends to a non zero value in this case, and the relation
between the asymptotic energy and the Edwards-Anderson parameter is the 
one verified in the TAP states, indicating equilibration within a metastable
state. For $q<q^*$ instead, the system looses the correlation with 
the initial state (and with $\rho$) and irrespectively of $q_{12}$,
falls into an aging state 
with asymptotic energy equal to $E_{dyn}(T)$ analogous
to the one discussed in \cite{cuku}, where the dynamics 
starts from a completely random initial condition. 
The situation can be understood analysing the energy of 
the initial state. For each value of $q_{12}$ this energy takes 
with probability one a fixed value $E(q_{12})$, which is a decreasing 
function of $q_{12}$ and equals $E_{eq}(T)$ for $q_{12}=1$. 
It is tempting at this point to interpret the initial states 
as {\it typical} states with that energy, i.e. as equilibrium 
states at a corresponding temperature $T(q_{12})$.
If this is true then for $T(q_{12})<T_{d}$ the typical initial configuration 
is in the basins of
metastable states that survive at temperature $T$ \cite{I,babumez,bfp},
although they are slightly deformed. The dynamics at temperature $T$
leads then to equilibrium in these states. 
For $T(q_{12})>T_{d}$ the typical initial configuration
belongs to the paramagnetic state, and aging has to be expected, this 
means that $T(q^*)=T_d$.
We have checked that this is indeed the right scenario using the techniques 
of \cite{I,babumez,bfp}, finding the interesting result that arbitrarily 
close to an equlibrium state at temperature $T$, there are states
which are of equilibrium at some other temperature. 

In this letter, we have studied  by a dynamical approach
the attraction basin of an equilibrium state at temperature 
$T_s < T < T_d$ (and established the correpondence with the 
static three-replicas potential). If we weight with the Boltzmann 
distribution we find wide basins of attraction. Almost all initial 
conditions with overlap larger than a threshold value $\bar{q}$,
are in the basin of attraction of the reference state. Conversely 
if we perform a white average we find zero size basins of attraction.
Starting close enough to the reference configuration,
 the system equilibrates in a TAP state close 
to the starting point, while if the initial overlap is larger 
than this threshold  the system ends up aging. 
This combination of facts indicates a highly
non trivial structure of the various states and basins of
attraction in configuration space.

{\bf Acknowledgements:} we thank A. Cavagna, I. Giardina and
M. Virasoro for useful discussions and comments; besides, we 
are most grateful to A. Cavagna and I. Giardina for providing us
with the values of the various parameters in the minima of the three-replicas
potential, thus allowing the quantitative comparison between effective 
potential and dynamics done in this letter.

\appendix

\section{Dynamical equations}

Denoting $f(q)=1/2 q^p$, and using the methods of 
\cite{I,babumez,houghton,kt}
one can show that the following dynamical equations are obeyed:
\bea
\mu(t) &=& \int_0^t ds\ ( f'(C(t,s))+f''(C(t,s))C(t,s) ) R(t,s)
+ \beta f'(Q(t))Q(t)
+ \beta \sum_{\alpha=1}^m f'(C_\alpha(t))C_\alpha(t) \nonumber \\
{\partial R(t,t') \over \partial t}&=& -\mu(t)  R(t,t')
+\int_{t'}^t ds  \ f''(C(t,s))R(t,s)R(s,t')
\nonumber \\
{\partial C(t,t') \over \partial t} &=&-\mu(t)C(t,t')
+ \int_0^{t'} ds \ f'(C(t,s))R(t',s)
+ \int_0^t ds\ f''(C(t,s))R(t,s)C(t',s)  \nonumber \\
& &+ \beta f'(Q(t))Q(t')
+ \beta  \sum_{\alpha=1}^m f'(C_\alpha(t))C_\alpha(t') \nonumber \\
\frac{d C_\alpha}{dt}  &=&-\mu(t)C_\alpha(t) +
\int_0^t ds \ f''(C(t,s))R(t,s)C_\alpha(s)
+ \beta f'(Q(t))q_{12}
+ \beta  \sum_{\beta=1}^m f'(C_\beta(t)) Q^{22}_{\alpha \beta}
 \nonumber \\
\frac{d Q_a}{dt}  &=&-\mu(t)Q_a(t) +
\int_0^t ds \ f''(C(t,s))R(t,s)Q_a(s)
+ \beta f'(Q_a(t))
+ \beta  \sum_{\alpha=1}^m f'(C_\alpha(t)) Q^{12}_{a,\alpha}
\label{A2}
\eea
where $Q^{12}$ and $Q^{22}$ are the matrices used in the 
two-replicas potential \cite{I}.
Using the 1-step RSB Ansatz, i.e.
$C_\alpha(t)=(C(t,0),C_1(t),C_0(t))$ with breaking point $x$,
with initial conditions
$C_1(0)=r_1$; $C_0(0)=r_0$, $Q(0)=q_{12}$, we can obtain the equations for
$C(t,t')$, $R(t,t')$, $C_1(t)$, $C_0(t)$, $Q(t)$ by expanding the sums in the
previous equations; for example:
\be
\sum_{\alpha=1}^m f'(C_\alpha(t))C_\alpha(t') =
f'(C(t,0))C(t',0)
+  (x-1) f'(C_1(t)) C_1(t')-  x f'(C_0(t))C_0(t')
\ee

\end{document}